\documentclass{article}
\usepackage{hiph-art}
\usepackage{graphicx}
\usepackage{amssymb}
\volnumber{00} \issuenumber{0} \edyear{2005}
\frompage{000} \topage{000}
\recrevdate{11 February 2005}
\def\rightmark{$\omega$ photoproduction in nuclei}
\def\leftmark{P.~Muehlich and U.~Mosel}
 
\title{In-Medium properties of the $\omega$ meson in photon induced nuclear reactions} 
\authors{ 
{P.~Muehlich$^{1,a}$ and U.~Mosel$^1$ %
\index{Muehlich, P.}
\index{Mosel, U.}
}\\[2.812mm]
{\normalsize
\hspace*{-8pt}$^{1}$ Institut fuer Theoretische Physik, Universitaet Giessen,\\ 
35392 Giessen, Germany\\[0.2ex] 
}}
 
\abstract{We discuss the possibility to study the in-medium changes of the properties of the $\omega$ meson in reactions on ordinary nuclei with elementary electromagnetic probes. We present a tree-level calculation of the elementary $\gamma p\rightarrow\omega p$ process which is extended to describe also the photoproduction of medium-modified $\omega$ mesons in nuclear matter. Using a semi-classical transport approach we obtain results for $e^+e^-$ and $\pi^0\gamma$ photoproduction off heavy nuclei in the invariant mass range of the $\rho$ and $\omega$ mesons. Both reactions are also studied experimentally and are presently being analyzed at accelerator facilities in Bonn and at Jefferson Lab. We show that the in-medium signals expected can be as large as those obtained in heavy-ion reactions.}

\makeindex
\begin{document}
 
\maketitle
%\setcounter{page}{1}

%%%%%%%%%%%%%%%%%%%%%%%%%%%%%%%%%%%%%%%%%%%%%%%%%%%%%%%%%%%%%%%%%%%%%%%%%%%%%%%%%%%%%%%%%%%%%%%%%%

\section{Introduction}\label{introduction}

One goal of current nuclear physics is to confirm that the strong interaction vacuum state indeed undergoes a phase transition towards the Wigner-Weyl realization of chiral symmetry, namely to a phase with a vanishing scalar quark condensate $\langle \bar qq\rangle$, as temperature and density increase. A unique tool to relate the quark and gluon condensates to actual observables is provided by the QCD sum rule approach that connects an expansion of the QCD current-current correlator in terms of quark and gluon degrees of freedom to an integral over the observable hadron spectrum (with the corresponding quantum numbers) \cite{Leupold:2001hj}. The measurement of the in-medium hadron spectrum can also give access to the density dependence of higher order quark condensates \cite{Zschocke:2002mp}.

Since the density modification of the condensates is expected to show traces of the phase transition already at rather moderate densities, a promising approach to learn about the modifications of hadrons in a strongly interacting environment is given by photoproduction experiments on nuclei. Apart from the fact that the initial state is almost purely governed by the electromagnetic interaction, i.~e. the interaction which is known best in elementary particle physics, such reactions have the advantage that the nuclear environment, i.~e. an ordinary nucleus, stays close to its ground state during the reaction and therefore provides a microscopic laboratory under well-defined conditions. Of special interest in this respect is the photoproduction of vector mesons in nuclei as they couple directly to virtual photons which again decay in a purely electromagnetic process to dilepton pairs, leaving the strongly interacting system untouched.

In the following we concentrate on the photoproduction of $\omega$ mesons and the possible observation of changes of its properties at finite baryon densities. A promising experiment looking for $e^+e^-$ pairs in the $\rho$ and $\omega$ mass range is presently being analyzed at JLab \cite{Jlab}. Another approach which focuses on the $\omega$ meson has been investigated at ELSA \cite{Trnka}. This second experiment looks for $\pi^0\gamma$ pairs produced in finite nuclei, which, however, are distorted by strong final state interactions of the pion. We have already shown in Refs. \cite{Muhlich:2003tj,Muhlich:2004cm}, that these FSI do not spoil the observation of the $\omega$ spectral density in the medium and can be even further suppressed by appropriate kinematic cuts. 

In the following we first present a model calculating the elementary $\gamma p\rightarrow\omega p$ process that is extended to describe the photoproduction of $\omega$ mesons in nuclear matter with arbitrary continuous mass spectra. In a second step we use the obtained cross section within a semi-classical transport model, describing the production, propagation and decay of vector mesons in nuclei within a coupled-channel treatment.

%%%%%%%%%%%%%%%%%%%%%%%%%%%%%%%%%%%%%%%%%%%%%%%%%%%%%%%%%%%%%%%%%%%%%%%%%%%%%%%%%%%%%%%%%%%%%%%%%%

\section{The model}\label{model}

\subsection{Photoproduction of $\omega$ mesons off nucleons}\label{nucleons}

One of the most successful descriptions of all existing data on $\omega$ photoproduction off the proton has been obtained with the coupled-channel model of Ref.~\cite{Penner:2002md}. There the importance of a coherent summation of amplitudes as well as the inclusion of rescattering effects and $s$- and $u$-channel resonance contributions has been pointed out. In a more recent work by the same group an even better describtion of the data could be obtained by including also higher-spin resonances \cite{Shklyar:2004ba}.

However, for our purpose the use of such an involved model is not possible and, anyway, goes beyond the scope of the present work. Hence, starting from the findings of the authors of Ref.~\cite{Penner:2002md}, we construct the tree-level cross section by taking over the most important contributions and refitting the open parameters, such as cutoffs and the $RN\omega$ coupling constants, in order to describe the experimental database.

\begin{figure}[htb]
\begin{center}
\vspace*{.5cm}
\includegraphics[scale=0.6]{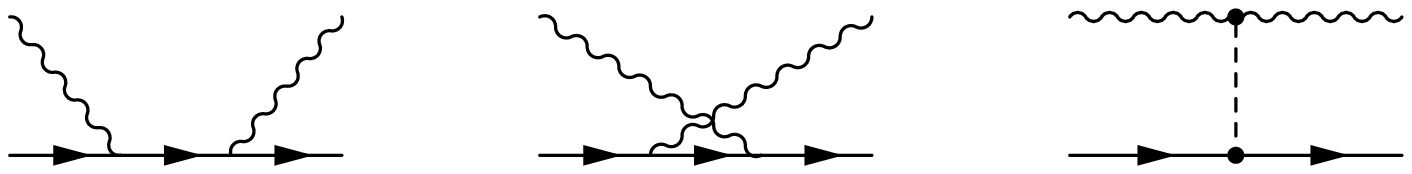}
\caption{Nucleon and pion exchange diagrams contributing to $\omega$ photoproduction}
\label{borndiag}
\end{center}
\end{figure}

\begin{figure}[htb]
\begin{center}
\vspace*{.5cm}
\includegraphics[scale=0.6]{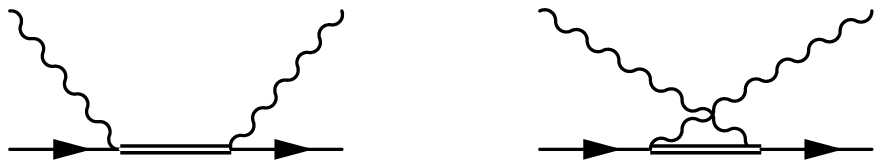}
\caption{$s$- and $u$-channel resonance contributions.}
\label{resdiag}
\end{center}
\end{figure}

\begin{figure}[!htb]
\begin{center}
\vspace*{.4cm}
\includegraphics[scale=1.1]{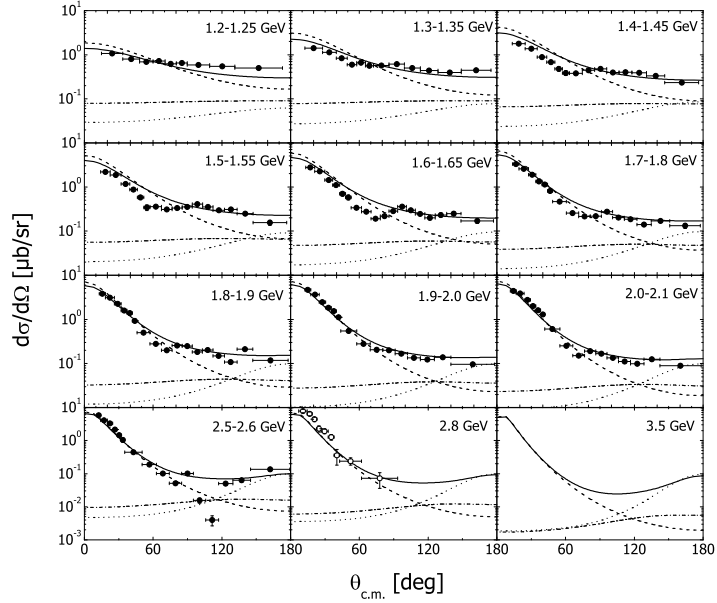}
\caption{Angular differential cross section for the process $\gamma p\rightarrow\omega p$. Data are taken from Ref.~\cite{Saphir} (solid symbols) and Ref.~\cite{Ballam:1972eq} (open symbols). Dashed lines: pion exchange contribution, dotted lines: resonance contributions, dash-dotted lines: $s$- and $u$-channel born diagrams, solid lines: coherent sum of all diagrams.}
\label{dsigdom}
\end{center}
\end{figure}

We include the $t$-channel $\pi$-exchange as well as the $s$- and $u$-channel nucleon diagrams depicted in Fig.~\ref{borndiag}. In order to ensure gauge invariance of the photoproduction amplitude we use the form-factor prescription of Ref.~\cite{Haberzettl:1998eq}. Since the authors of Ref.~\cite{Penner:2002md} found an important contribution of the $P_{11}(1710)$ nucleon resonance, we also include the $s$- and $u$-channel resonance contributions, see Fig.~\ref{resdiag}. At the $RN\omega$ vertices we apply the same form-factors as on the corresponding nucleon vertices. The results for the angular differential cross section, that is obtained by a coherent summation of the individual contributions, are shown in Fig.~\ref{dsigdom}. A satisfactory description of the data can be achieved.

%%%%%%%%%%%%%%%%%%%%%%%%%%%%%%%%%%%%%%%%%%%%%%%%%%%%%%%%%%%%%%%%%%%%%%%%%%%%%%%%%%%%%%%%%%%%%%

\subsection{The $\omega$ meson in nuclear matter}\label{matter}

\begin{figure}[htb]
\begin{center}
\vspace*{.4cm}
\includegraphics[scale=0.7]{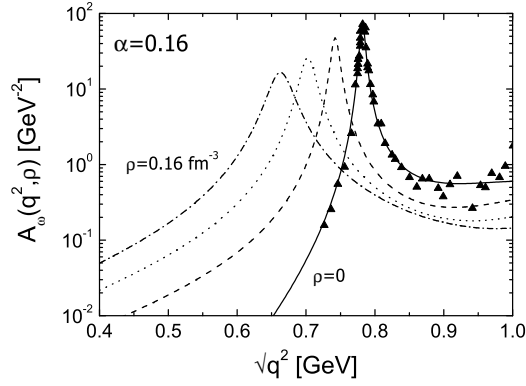}
\caption{Parametrization of the $\omega$ spectral function in nuclear matter for $\omega$ mesons at rest.}
\label{spectral}
\end{center}
\end{figure}

Various models predict a sizable change of the mass and width of the $\omega$ meson as soon as it is put in a strongly interacting environment (see \cite{Muhlich:2003tj} for references). All these informations are encoded in the spectral function $\mathcal{A}_{\omega}$ which is the imaginary part of the $\omega$ in-medium propagator:
\begin{eqnarray}
\mathcal{A}_{\omega}(q)=-\frac{1}{\pi}\mathrm{Im}\frac{1}{q^2-m_{\omega}^2-\Pi(q)},
\end{eqnarray}
with the (spin-averaged) in-medium self energy $\Pi(q)$. Within the semi-classical transport approach the imaginary part of this self energy is given as a sum of the vacuum self energy and a second term containing the effects of collisional broadening:
\begin{eqnarray}
\Pi(q^2,|{\bf q}|)=2EV\simeq -2\alpha m_{\omega}^2\frac{\rho}{\rho_0}-i\left(\sqrt{q^2}\Gamma_{\mathrm{vac}}(q^2)+|{\bf q}|\rho\sigma_{\omega N}\right),
\label{self}
\end{eqnarray}
where $\sigma_{\omega N}$ stands for the $\omega N$ total cross section. The real part in Eq.~(\ref{self}) has been parameterized by a momentum independent mass shift linear in the nuclear density with the mass shift parameter $\alpha=-\Delta m/m$. The such parameterized spectral function is shown in Fig.~\ref{spectral}, using a value of $\alpha=0.16$ according to Ref.~\cite{Klingl:1998zj}.

%%%%%%%%%%%%%%%%%%%%%%%%%%%%%%%%%%%%%%%%%%%%%%%%%%%%%%%%%%%%%%%%%%%%%%%%%%%%%%%%%%%%%%%%%%%%%%

\subsection{$\omega$ photoproduction in matter}\label{nuclei}

The $\omega$ photoproduction cross section in nuclear matter is obtained by allowing the $\omega$ to take arbitrary masses, leading to an additional integration over the 0-$th$ component of the $\omega$ four-momentum in the phase-space integral:
\begin{eqnarray}
\sigma(E,\rho)=\frac{1}{2E_{\gamma}2E_N}\frac{1}{|{\bf v}_{\gamma}-{\bf v}_N|}\int \frac{d^3p_{N'}}{(2\pi)^32E_{N'}}\frac{dq^0d^3q}{(2\pi)^3}|\mathcal{M}|^2\nonumber\\
\qquad\qquad\times(2\pi)^4\delta^4(p_{\gamma}+p_N-p_{N'}-q)\mathcal{A}_{\omega}(q,\rho({\bf r})),
\end{eqnarray}
where $\mathcal{M}$ is the photoproduction amplitude discussed in the previous section, depending in a well defined way on the $\omega$ meson mass.

\begin{figure}[htb]
\begin{center}
\vspace*{.3cm}
\includegraphics[scale=0.7]{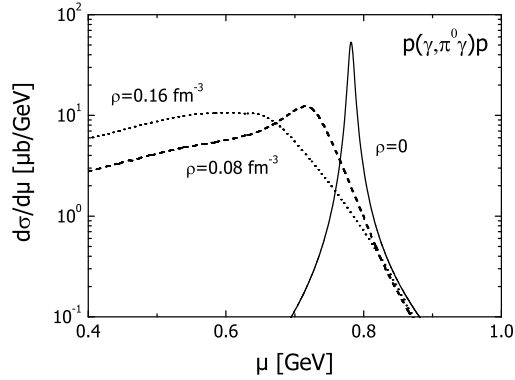}
\caption{Mass differential cross section for the process $p(\gamma,\pi^0\gamma')p$ in nuclear matter at various densities.}
\label{msigma}
\vspace*{-.25cm}
\end{center}
\end{figure}

In Fig.~\ref{msigma} we have plotted the mass differential cross section for the process $\gamma N\rightarrow\omega N'\rightarrow\pi^0\gamma N'$ at various nuclear densities. Due to the shift of spectral strength to lower $\omega$ masses we find a tremendous enhancement of the cross section for low-mass $\pi^0\gamma$ pairs. This behavior has also to do with the inclusion of the $P_{11}$ resonance terms in the amplitude. Since this very broad resonance sits right at the $\omega N$ threshold, about half of its spectral strength does not contribute to the photoproduction of on-shell $\omega$ mesons. As soon as the $\omega$ is shifted to lower masses, more of the spectral strength of the $P_{11}$ falls into the kinematically allowed region, hence leading to an enhancement of the $\omega$ photoproduction cross section at finite densities.

%%%%%%%%%%%%%%%%%%%%%%%%%%%%%%%%%%%%%%%%%%%%%%%%%%%%%%%%%%%%%%%%%%%%%%%%%%%%%%%%%%%%%%%%%%%%%%

\subsection{Photoproduction from finite nuclei}\label{transport}

Photoproduction reactions in finite nuclei we model employing a semi-classical Boltzmann-Uehling-Uhlenbeck (BUU) transport approach. Due to the smallness of the electromagnetic coupling constant $\alpha_{\mathrm{em}}\approx 1/137$, the interaction of the incoming photon can be treated perturbatively, leading to a two-step description of the photon-nucleus reaction: In the first step the incoming (and potentially shadowed, see e.~g. \cite{Falter:2002jc}) photon interacts with a single nucleon of the target nucleus and produces one or several particles. In the second step these particles are propagated through the nuclear medium, including elastic and inelastic scattering, absorption and decay. The correct behavior of the spectral functions is also taken care of by means of a phenomenological off-shell potential, which shifts the particle masses back to their vacuum values as the local density vanishes. This method and more details of the BUU model are described in Refs.~\cite{Muhlich:2002tu,Effenberger:1999ay} and references therein.

%%%%%%%%%%%%%%%%%%%%%%%%%%%%%%%%%%%%%%%%%%%%%%%%%%%%%%%%%%%%%%%%%%%%%%%%%%%%%%%%%%%%%%%%%%%%%%

\section{Results}

\subsection{Dilepton production}

The observation of the dilepton invariant mass spectrum has the advantage that neither the initial nor the final state is distorted by initial (final) state interactions. In the considered energy regime the most important contributions to $e^+e^-$ production come from the direct decay of the light vector mesons $\rho,~\omega$ and $\phi$ via $V\rightarrow\gamma^*\rightarrow e^+e^-$ and from the Daltiz decays $\Delta\rightarrow Ne^+e^-, \eta\rightarrow\gamma e^+e^-, \omega\rightarrow\pi^0 e^+e^-$ and $\pi^0\rightarrow\gamma e^+e^-$. For details see Ref.~\cite{Effenberger:1999ay}.

\begin{figure}[htb]
\begin{center}
\vspace*{.3cm}
\includegraphics[scale=0.7]{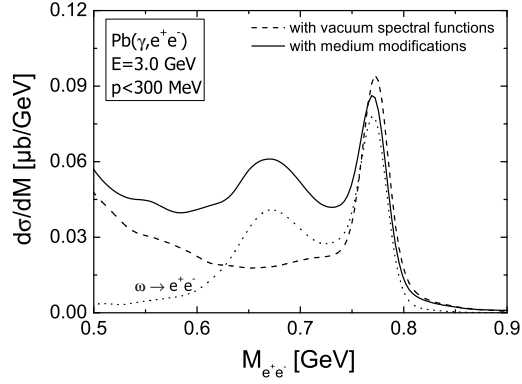}
\caption{Dilepton invariant mass spectrum from a lead target with a 3-momentum cutoff of 300 MeV.}
\label{dilepton}
\end{center}
\end{figure}

The sensitivity of the $e^+e^-$ mass spectrum to the in-medium decays of the vector mesons can even be enhanced by reducing the decay length $L_V=p_V/(m_V\Gamma_V)$ of the vector meson under consideration. This can be done by applying a cutoff on the three-momentum of the detected $e^+e^-$ pair. In Fig.~\ref{dilepton} we show a typical invariant mass spectrum obtained from a lead target. The cross section including the $\omega$ medium modifications according to Eq.~(\ref{self}) even shows a shifted and broadened $\omega$ peak that allows for a quantitative extraction of the modification of the $\omega$ mass in nuclei.

%%%%%%%%%%%%%%%%%%%%%%%%%%%%%%%%%%%%%%%%%%%%%%%%%%%%%%%%%%%%%%%%%%%%%%%%%%%%%%%%%%%%%%%%%%%%%%

\subsection{$\pi^0\gamma$ production}

\begin{figure}[htb]
\begin{center}
\vspace*{.3cm}
\includegraphics[scale=0.7]{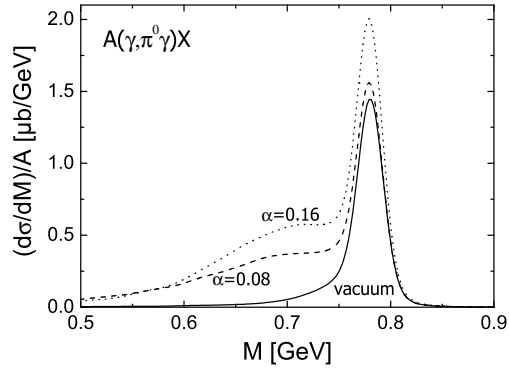}
\caption{$\pi^0\gamma$ invariant mass spectrum from a $Nb$ target for different mass shift parameters.}
\label{cbtaps}
\end{center}
\end{figure}

The semi-hadronic final state $\pi^0\gamma$ is a priori less sensitive to any in-medium modification as the mean free path of the pion in nuclear matter is quite small $\lambda\sim1~\mathrm{fm}$. Nevertheless, distorted $\pi^0\gamma$ pairs in the final state do not contribute strongly in the invariant mass range of interest (0.6-0.9 GeV) and can even be further suppressed by cutting on the kinetic energy of the detected pion $T_{\pi}\geq 150~\mathrm{MeV}$, for details see Ref.~\cite{Muhlich:2003tj}. In Fig.~\ref{cbtaps} we show the $\pi^0\gamma$ invariant mass spectrum from a Niobium target using different values for the $\omega$ mass shift parameter $\alpha$. The $\omega$ vacuum peak (already broadened from the finite detector resolution of 25 MeV) is always clearly visible due to $\omega$ decays outside the nucleus. The shift of spectral strength to lower invariant masses causes a clearly visible shoulder on the left side of the $\omega$ peak. One can also see that the integral under the curve increases with higher values for the mass shift parameter. This is due to the previously discussed increase of phase-space in the $\omega$ photoproduction process.

%%%%%%%%%%%%%%%%%%%%%%%%%%%%%%%%%%%%%%%%%%%%%%%%%%%%%%%%%%%%%%%%%%%%%%%%%%%%%%%%%%%%%%%%%%%%%%
 
\section{Summary}

We have shown the sensitivity of both the $e^+e^-$ and the $\pi^0\gamma$ final state to the in-medium properties of the $\omega$ meson in photoproduction reactions from heavy nuclei. Despite the strong final state interactions in the case of the semi-hadronic final state, an observation of modified in-medium properties of the $\omega$ meson has proven to be possible, provided the in-medium changes are sizeable and also more complex as a pure broadening of the peak. The non-hadronic final state $e^+e^-$ could even give access to quantitative information about a possible mass shift that is relevant also to the understanding of high density phenomena in ultra relativistic heavy ion collisions.

%%%%%%%%%%%%%%%%%%%%%%%%%%%%%%%%%%%%%%%%%%%%%%%%%%%%%%%%%%%%%%%%%%%%%%%%%%%%%%%

%\section*{Appendix}\label{app}
%\begin{eqnarray}
%\mathcal{L}_{\gamma NN} & = & -e\bar %u_N\left(\gamma_{\mu}-\frac{\kappa_N}{2m_N}\sigma_{\mu\nu}\partial^{\nu}\right)A^{\mu}u_N\\
%\mathcal{L}_{\omega NN} & = & -g_{\omega}\bar %u_N\left(\gamma_{\mu}-\frac{\kappa_{\omega}}{2m_N}\sigma_{\mu\nu}\partial^{\nu}\right)\omega^{\mu}u_N\\
%\mathcal{L}_{\omega\pi\gamma} & = & %\frac{f_{\omega\pi\gamma}}{m_{\omega}}\epsilon_{\mu\nu\alpha\beta}\partial^{\mu}A^{\nu}\partial^{\alpha}\omega^{\beta}\pi_0\\
%\mathcal{L}_{\pi NN} & = & -ig_{\pi NN}\bar u_N\gamma^5u_N\pi_0
%\end{eqnarray}
%\begin{eqnarray}
%F_t(q^2)=\left(\frac{\Lambda_{\omega\pi\gamma}^2-q^2}{\Lambda_{\omega\pi\gamma}^2-m_{\pi}^2}\right)\left(\frac{\Lambda_{\pi %NN}^2-q^2}{\Lambda_{\pi NN}^2-m_{\pi}^2}\right)
%\end{eqnarray}
%\begin{eqnarray}
%\mathcal{L}_{RN\gamma}=e\frac{g_1}{2m_N}\bar u_R\sigma_{\mu\nu}\partial^{\nu}_Au_NA^{\mu}
%\end{eqnarray}
%\begin{eqnarray}
%\mathcal{L}_{RN\omega}=-\bar %u_R\left(g_1\gamma_{\mu}-\frac{g_2}{2m_N}\sigma_{\mu\nu}\partial_{\omega}^{\nu}\right)u_N\omega^{\mu}
%\end{eqnarray}
%\begin{eqnarray}
%F(q^2,m^2)=\frac{\Lambda^4}{\Lambda^4+(q^2-m^2)^2}
%\end{eqnarray}
%\begin{eqnarray}
%F(q^2,m^2)=\alpha F_s(q^2,m^2)+(1-\alpha)F_u(q^2,m^2)
%\end{eqnarray}

%%%%%%%%%%%%%%%%%%%%%%%%%%%%%%%%%%%%%%%%%%%%%%%%%%%%%%%%%%%%%%%%%%%%%%%%%%%%%%%

\section*{Acknowledgment}

This work has been supported by the Deutsche Forschungsgemeinschaft (DFG). We gratefully acknowledge support by the Frankfurt Center for Scientific Computing (CSC).

%%%%%%%%%%%%%%%%%%%%%%%%%%%%%%%%%%%%%%%%%%%%%%%%%%%%%%%%%%%%%%%%%%%%%%%%%%%%%%%
 
\section*{Notes}
\begin{notes}
\item[a] 
E-mail: muehlich@theo.physik.uni-giessen.de
\end{notes}

%%%%%%%%%%%%%%%%%%%%%%%%%%%%%%%%%%%%%%%%%%%%%%%%%%%%%%%%%%%%%%%%%%%%%%%%%%%%%%%

%%%%%%%%%%%%%%%%%%%%%%%%%%%%%%%%%%%%%%%%%%%%%%%%%%%%%%%%%%%%%%%%%%%%%%%%%%%%%%%

\vfill\eject

\begin{thebibliography}{99}  

\bibitem{Leupold:2001hj}
S.~Leupold,
%``QCD sum rule analysis for light vector and axial-vector mesons in  vacuum
%and nuclear matter,''
{\it Phys.\ Rev.}\ C {\bf 64} (2001) 015202.
%[arXiv:nucl-th/0101013].
%%CITATION = NUCL-TH 0101013;%%

\bibitem{Zschocke:2002mp}
S.~Zschocke, O.~P.~Pavlenko and B.~Kampfer,
%``In-medium spectral change of omega mesons as a probe of QCD four-quark
%condensate,''
{\it Phys.\ Lett.}\ B {\bf 562} (2003) 57.
%[arXiv:hep-ph/0212201].
%%CITATION = HEP-PH 0212201;%%

\bibitem{Jlab}
S.~Boiarinov et al., {\it Jefferson Lab Proposal} E-01-112 (2001), unpublished;
C.~Tur, D.~Weygand, M.~Wood and C.~Djalali, {\it priv. communication} (2004).

\bibitem{Trnka}
D.~Trnka et al., (CB-Taps collaboration), in preparation.

\bibitem{Muhlich:2003tj}
P.~Muhlich, T.~Falter and U.~Mosel,
%``Inclusive omega photoproduction off nuclei,''
{\it Eur.\ Phys.\ J.}\ A {\bf 20} (2004) 499.
%[arXiv:nucl-th/0310067].
%%CITATION = NUCL-TH 0310067;%%

\bibitem{Muhlich:2004cm}
P.~Muhlich, T.~Falter and U.~Mosel,
%``In-medium properties of the omega meson through omega photoproduction in
%nuclei,''
arXiv:nucl-th/0402039.
%%CITATION = NUCL-TH 0402039;%%

\bibitem{Penner:2002md}
G.~Penner and U.~Mosel,
%``Vector meson production and nucleon resonance analysis in a coupled  channel
%approach for energies m(N) < s**(1/2) < 2-GeV. II: Photon  induced results,''
{\it Phys.\ Rev.}\ C {\bf 66} (2002) 055212.
%[arXiv:nucl-th/0207069].
%%CITATION = NUCL-TH 0207069;%%

\bibitem{Shklyar:2004ba}
V.~Shklyar, H.~Lenske, U.~Mosel and G.~Penner,
%``Coupled-channel analysis of the omega-meson production in pi N and gamma N
%reactions for cm energies up to 2-GeV,''
arXiv:nucl-th/0412029.
%%CITATION = NUCL-TH 0412029;%%

\bibitem{Haberzettl:1998eq}
H.~Haberzettl, C.~Bennhold, T.~Mart, T.~Feuster,
%``Gauge-invariant tree-level photoproduction amplitudes with form  factors,''
{\it Phys.\ Rev.}\ C {\bf 58} (1998) 40.
%[arXiv:nucl-th/9804051].
%%CITATION = NUCL-TH 9804051;%%

\bibitem{Saphir}
J.~Barth {\it et al.}, 
{\it Eur.~Phys.~J.} A {\bf 18} (2003) 117.

\bibitem{Ballam:1972eq}
J.~Ballam {\it et al.},
%``Vector Meson Production By Polarized Photons At 2.8-Gev, 4.7-Gev, And
%9.3-Gev,''
{\it Phys.\ Rev.}\ D {\bf 7} (1973) 3150.
%%CITATION = PHRVA,D7,3150;%%

\bibitem{Klingl:1998zj}
F.~Klingl, T.~Waas and W.~Weise,
%``Nuclear bound states of omega mesons,''
{\it Nucl.\ Phys.}\ A {\bf 650} (1999) 299.
%[arXiv:hep-ph/9810312].
%%CITATION = HEP-PH 9810312;%%

\bibitem{Falter:2002jc}
T.~Falter and U.~Mosel,
%``Hadron formation in high energy photonuclear reactions,''
{\it Phys.\ Rev.}\ C {\bf 66} (2002) 024608.
%[arXiv:nucl-th/0203052].
%%CITATION = NUCL-TH 0203052;%%

\bibitem{Muhlich:2002tu}
P.~Muhlich, T.~Falter, C.~Greiner, J.~Lehr, M.~Post and U.~Mosel,
%``Photoproduction of Phi mesons from nuclei,''
{\it Phys.\ Rev.}\ C {\bf 67}, 024605 (2003).
%[arXiv:nucl-th/0210079].
%%CITATION = NUCL-TH 0210079;%%

\bibitem{Effenberger:1999ay}
M.~Effenberger, E.~L.~Bratkovskaya, U.~Mosel,
%``e+ e- pair production from gamma A reactions,''
{\it Phys.\ Rev.}\ C {\bf 60}, (1999) 044614.
    
\end{thebibliography}
\end{document}